\documentclass[prl,twocolumn,notitlepage,showpacs,amsmath,amstex,amssymb,citeautoscript,longbibliography,superscriptaddress]{revtex4-1}
\pdfoutput=1

\usepackage{natbib}
\usepackage[english]{babel}
\usepackage{letltxmacro}
\usepackage{latexsym}
\LetLtxMacro{\ORIGselectlanguage}{\selectlanguage}
\makeatletter
\DeclareRobustCommand{\selectlanguage}[1]{%
  \@ifundefined{alias@\string#1}
    {\ORIGselectlanguage{#1}}
    {\begingroup\edef\x{\endgroup
       \noexpand\ORIGselectlanguage{\@nameuse{alias@#1}}}\x}%
}
\newcommand{\definelanguagealias}[2]{%
  \@namedef{alias@#1}{#2}%
}
\makeatother
\definelanguagealias{en}{english}
\definelanguagealias{English}{english}
\usepackage{graphicx}
\usepackage{amsmath}
\usepackage{mathtools}
\usepackage{amsfonts}
\usepackage{amssymb,bbding}
\usepackage{bm}
\usepackage{color}
\usepackage[percent]{overpic}
\usepackage{soul} 
\usepackage{amssymb}
\usepackage{wasysym}
\usepackage{dsfont}
\usepackage{float}
\usepackage{braket}
\usepackage{subfiles}
\usepackage{epstopdf}

\usepackage[normalem]{ulem}

\usepackage{tikz}
\usetikzlibrary{positioning}

\usepackage{blkarray}
\usepackage{multirow}

\usepackage{hyperref}
\hypersetup{
    bookmarks=false,         
    unicode=false,          
    pdftoolbar=false,        
    pdfmenubar=true,        
    pdffitwindow=false,     
    pdfstartview={FitH},    
    pdftitle={Analytically solvable renormalization group for the many-body localization transition},    
    pdfauthor={A. Goremykina, R. Vasseur, M. Serbyn},     
    pdfsubject={},   
    pdfcreator={},   
    pdfproducer={}, 
    pdfkeywords={MBL} {MBL transition}, 
    pdfnewwindow=true,      
    colorlinks=true,       
    linkcolor=black,          
    citecolor=blue,        
    filecolor=magenta,      
    urlcolor=blue           
}

\newcommand{\be}{\begin{equation}}
\newcommand{\ee}{\end{equation}}
\newcommand{\bea}{\begin{eqnarray}}
\newcommand{\eea}{\end{eqnarray}}

\newcommand{\bt}{\kappa}
\newcommand{\rt}{\gamma}

\newcommand{\QIT}{Q_*^{I,T}}
\newcommand{\QT}{Q_*^{T}}
\newcommand{\QI}{Q_*^{I}}
\newcommand{\fIT}{f^{I,T}}
\newcommand{\fT}{f^{T}}
\newcommand{\fI}{f^{I}}
\newcommand{\OIT}{\hat{O}_{I,T}}
\newcommand{\calO}{{\cal O}}

\begin{document}

\title{Analytically solvable renormalization group for the many-body localization transition}

\author{Anna Goremykina}
\affiliation{D\'epartement de Physique Th\'eorique, Universit\'e de Gen\`eve, CH-1211 Gen\`eve 4, Switzerland}
\affiliation{IST Austria, Am Campus 1, 3400 Klosterneuburg, Austria}

\author{Romain Vasseur}
\affiliation{Department of Physics, University of Massachusetts, Amherst, Massachusetts 01003, USA}

\author{Maksym Serbyn }
\affiliation{IST Austria, Am Campus 1, 3400 Klosterneuburg, Austria}

\date{\today}
\begin{abstract}
We introduce a simple, exactly solvable strong-randomness renormalization group (RG) model for the many-body localization (MBL) transition in one dimension. Our approach relies on a family of RG flows parametrized by the asymmetry between thermal and localized phases. We identify the physical MBL transition in the limit of maximal asymmetry, reflecting the instability of MBL against rare thermal inclusions. We find a critical point that is localized with power-law distributed thermal inclusions. The typical size of critical inclusions remains finite at the transition, while the average size is logarithmically diverging. We propose a two-parameter scaling theory for the many-body localization transition that falls into the Kosterlitz-Thouless universality class, with the MBL phase corresponding to a stable line of fixed points with multifractal behavior. 
\end{abstract}
\maketitle

{\it Introduction.}--- The many-body localization transition (MBLT) separates many-body localized (MBL) and ergodic dynamical phases in isolated quantum systems~\cite{Anderson80,Levitov97,Mirlin05,Basko06,PalHuse,DavidRahulReview,EhudRonenReview,JoelRomainReview,2017arXiv171103145A,AbaninRev}. On the ergodic or thermal side of this transition, the system exchanges energy and information efficiently between its parts, thus quickly loosing its quantum nature. This corresponds to an extensive amount of quantum entanglement in many-body eigenstates. In contrast, the MBL phase is non-ergodic and avoids thermalization by means of an extensive number of local conserved quantities~\cite{Serbyn13-1,Huse13, ImbriePRL}. The high energy eigenstates of MBL systems have low area-law entanglement~\cite{Serbyn13-1,Bauer13} and allow one to encode quantum information even at long times~\cite{Bahri,Serbyn_14_Deer}. 

Although MBLT in one dimension is a subject of intense theoretical~\cite{PalHuse,PhysRevB.90.224203,VHA,PVP,DVP,huse,TMRshort,TMR,Serbyn15,Alet14,KhemaniEnt} and experimental studies~\cite{Bloch15,Bloch16-2,Bloch16G}, many aspects of this phase transition remain poorly understood or debated. Numerical studies 
are limited to very small system sizes and are believed to suffer from finite size effects~\cite{2015arXiv150904285C}. On the other hand, the pioneering strong disorder renormalization group (RG) approaches by~\textcite{VHA} and~\textcite{PVP} evaded analytical solutions and relied on numerical simulations of simplified RG rules. The recent RG approach by~\textcite{TMRshort,TMR} allowed for ``mean-field'' approximate solution, however resulting in unphysical exponents.

Recently~\textcite{huse} introduced an exactly solvable RG for the MBLT. However, this RG has an inherent unphysical symmetry between MBL and thermal phases. Typically, the ergodic behavior and tendency to form resonances is very strong in quantum systems.  On this basis, one expects that even a sparse network of resonances~\cite{PVP,KhemaniEnt} suffices for delocalization, and thus the critical point between MBL and ergodic phases should be more similar to the localized phase~\cite{VHA,TMRshort}. These expectations are confirmed by numerical studies~\cite{KhemaniEnt} and also earlier RG studies~\cite{VHA,PVP,DVP,TMR}.

In this work we present an analytically solvable family of strong-randomness RGs, which can be viewed as a deformation of the RG studied previously in~\cite{huse}.  The deformation is parametrized by $\alpha\leq 1$ that sets the asymmetry between MBL and thermal phase at the transition. We calculate the correlation length exponent $\nu$ and fractal dimensions for generic values of $\alpha$. Upon decreasing value of $\alpha$ we observe that $\nu$ diverges,  the critical point looks progressively more insulating, and the distribution of thermal blocks tends to a scale-invariant power-law shape. We identify the physical MBLT with the limit of maximal asymmetry $\alpha \to 0$ when critical point is localized with a probability one. By analytically continuing our RG equations to the case $\alpha\to 0$, we find that the MBL phase corresponds to a line of fixed points with the length of thermal inclusions distributed according to the power-law distribution $\rho_T(\ell) \sim 1/\ell^{2+\kappa}$ for large $\ell$ with $\kappa \geq 0$. The transition to the thermal phase occurs for the critical value $\kappa_c=0$, when the average size of thermal inclusions diverges (while typical thermal puddles remain finite). We find that the thermal inclusions are renormalized by the surrounding MBL phase upon coarse graining, leading us to a simple two-parameter RG theory in the Kosterlitz-Thouless universality class~\cite{KT}. This implies that the correlation length is diverging exponentially at the transition, in sharp contrast with previous predictions.

{\it Two-parameter family of RGs.}---To develop a theory of the MBLT, we adopt a coarse-grained picture~\cite{PVP,VHA,huse,DVP,TMRshort} and assume that at some intermediate length scale the critical system can be viewed as a set of thermal (ergodic) and insulating~(MBL) regions. Starting from this length scale, we build the RG description to account for the competition between ergodic regions that tend to hybridize the nearby insulators and MBL clusters that absorb thermal regions and prevent resonances. 

\begin{figure}[t]
\begin{center}
\includegraphics[width=0.99\columnwidth]{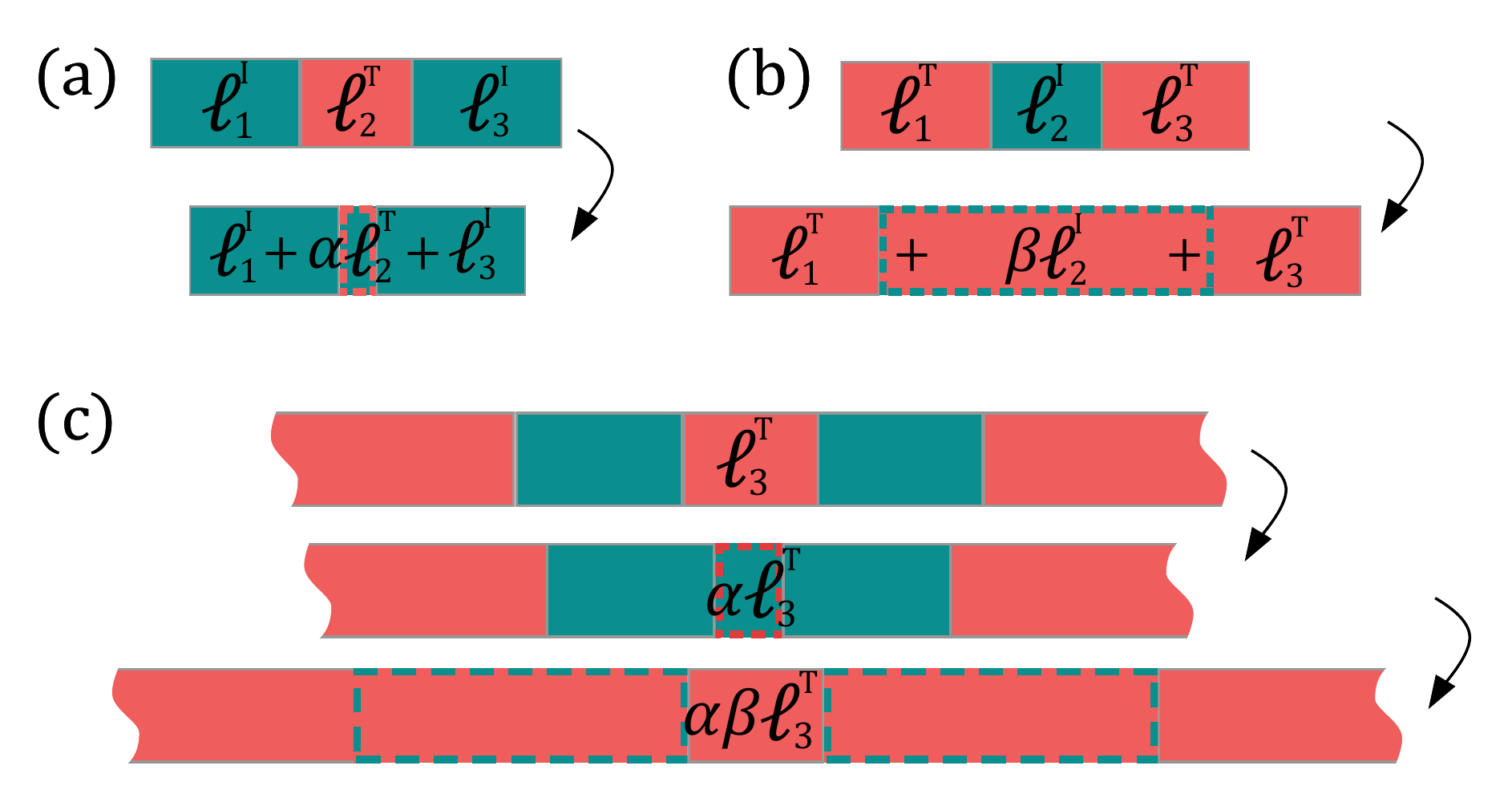}
\caption{ \label{Fig:rules}
Illustration of RG rules for the decimation of thermal (a) and insulating segments (b). (c) The length of the central thermal block $\ell^T_2$ is recovered with unit prefactor after two decimation steps if $\alpha\beta=1$.}
\end{center}
\end{figure}

Aiming for a simple description~\cite{huse}, we assume that each region can be characterized by a single parameter $\ell$, which we refer to as ``length''. At  each RG step one removes the shortest thermal (insulating) segment by merging it with adjacent  insulating (thermal) regions; see Fig.~\ref{Fig:rules}. The length of a new region reads 
\begin{equation}\label{Eq:RG_rules}
\ell^I_\text{new} =  \ell^{I}_{n-1} + \alpha \ell^{T}_n + \ell^{I}_{n+1},
\
\ell^T_\text{new} =  \ell^{T}_{n-1} + \beta \ell^{I}_n + \ell^{T}_{n+1},
\end{equation}
where the length of the decimated segment is multiplied by a parameter $\alpha$ if it is thermal, and by $\beta$ if it is insulating.  

Equations~(\ref{Eq:RG_rules}) describe a two-parameter family of RGs, which reduces to (over)simplified RG in Ref.~\cite{huse} for  symmetric point $\alpha=\beta=1$. We seek a deformation away from this point that makes the critical point more MBL-like. Intuitively, such asymmetry reflects the very strong tendency of quantum chaotic systems to develop entanglement and form resonances. Hence, even a small fraction of thermal blocks should suffice to drive the transition. Such deformation can be achieved by taking $\alpha\ll1\ll\beta$ so that the insulating segments do not increase much in size when a thermal block is decimated. On the other hand, when two thermal blocks absorb an insulating segment, the resulting thermal region has a significantly larger length, see Fig.~\ref{Fig:rules}(b), hence being less likely to be decimated again.

 The rules in Eq.~(\ref{Eq:RG_rules}) with  $\alpha\ll1\ll\beta$ are physically motivated if we interpret the length $\ell$ as setting the hybridization time $\tau$ through the corresponding segment. For insulating blocks it is natural to assume the time to be exponentially large in $\ell^I$, $\tau^I \propto \exp({\ell^I/\xi_0})$, where $\xi_0$ is the (bare) localization length. In contrast, for thermal segments such time is expected to scale as $\tau^T\propto \ell^T$.  When a thermal segment is decimated, its  contribution to the hybridization rate of a new segment is negligible, motivating a small value of $\alpha$. Similarly, the large value of $\beta$ mimics the dominant contribution of the $I$ segment to the hybridization time of a $TIT$ block; see Supplemental Material~\cite{SOM} for more details. Moreover, this limit of variables will be justified in the following by the condition $\alpha \beta=1$ and by the absence of fractality of the insulating regions at criticality when $\alpha = 0$.

{\it Generalized length-preserving line $\alpha\beta=1$.}---We can obtain an additional relation between the parameters $\alpha,\beta$ by imposing that the contribution of each segment does not depend on its previous history in the RG. For instance, Fig.~\ref{Fig:rules}(c) shows a microscopic thermal segment of length $\ell^T_3$ that first was absorbed into an insulating segment, but later becomes again part of a thermal region. Requiring that this segment contributes by the amount $\ell^T_3$ to the effective length of the final thermal region, we obtain the condition $\alpha\beta=1$. When $\alpha\beta=1$, one can define a generalized total length, $\ell_\text{tot} = \sum_n (\alpha\ell^T_n +  \ell^I_n)$, that is preserved along the RG flow. If $\alpha\to 0$, it results in the conservation of the total length of insulating regions, which guarantees the correct scaling between the tunneling time and the total length when flowing into the localized phase.
In what follows we restrict to the line $\alpha\beta=1$, using value of $\alpha<1$ as a control parameter. Critical behavior for generic values of $\alpha, \beta$ will be reported elsewhere~\cite{wetobe}.

\begin{figure*}[t]
\includegraphics[width=1.99\columnwidth]{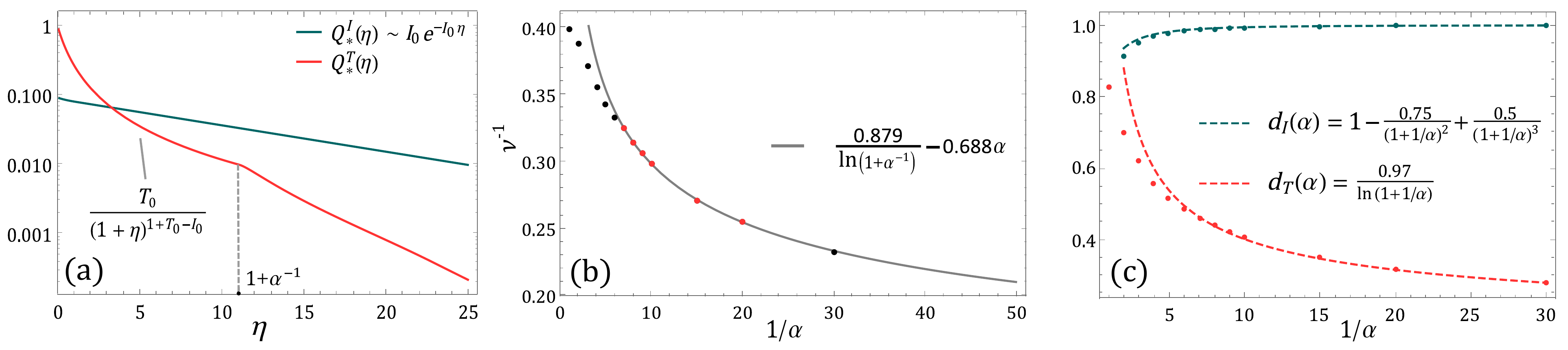} 
\caption{\label{Fig:exponent}
(a) For small $\alpha=\beta^{-1}=1/10$ the fixed point distribution for thermal $Q^{\ast}_T(\eta)$ blocks (red line) behaves as a power-law for $\eta\leq \alpha^{-1}+1$, and decays exponentially for larger values of $\eta$. Insulating blocks are approximately distributed exponentially for all $\eta$ (blue line). 
(b) Slow decay of inverse critical exponent $\nu^{-1}$ with $\alpha^{-1}$ is well approximated by the analytical asymptotic. The dots marked with red have been used for the extrapolation of $\nu^{-1}(\alpha)$ for smaller values of $\alpha$. Value of $\nu^{-1}(1/30)$ collapses well onto the numerical fit. 
(c) Fractal dimension $d_I$ of insulating regions rapidly approaches one when $\alpha\to 0$, whereas fractal dimension of thermal inclusions slowly decays to zero.}
\end{figure*}

{\it Flow equations.}---In order to describe the critical point, we derive RG flow equations for distributions of lengths of MBL and thermal segments~\cite{dasgupta_ma, ma_dasgupta_hu,fisher}. At each step of the RG, the smallest block of length $\Gamma \equiv \min \ell_n$ is decimated according to the rules~(\ref{Fig:rules}).  Let $\rho^{I, T}_{\Gamma}(\ell)$ be the distributions of insulating and thermal block lengths respectively, with cutoff~$\Gamma$. It is convenient to define the rescaled dimensionless length $\eta =({\ell-\Gamma})/{\Gamma}$ and associated probability distributions $\rho^{I,T}_\Gamma(\ell) = ({1}/{\Gamma}) Q^{I,T}_{\Gamma}(\eta)$.
The RG equations which describe the flow of these rescaled probability distributions with the cutoff~$\Gamma$ read as~\cite{rb, fisher}: 
\begin{widetext}
\begin{subequations}
\label{RG_eqs}
\begin{eqnarray}\label{RG_eqsI}
\frac{\partial Q_\Gamma^{I}(\eta)}{\partial \ln \Gamma}
&=&\partial_\eta \big[(1 + \eta)Q_\Gamma^{I}(\eta)\big]
+Q_\Gamma^{I}(\eta) [Q_\Gamma^{I}(0)-Q_\Gamma^{T}(0)]
+Q_\Gamma^{T}(0) \theta(\eta -\alpha-1)
\int_0^{\eta -\alpha- 1}\! \! \! \! \! \!\! \! d\eta'Q^{I}_\Gamma(\eta')Q^{I}_\Gamma(\eta - \eta' -\alpha- 1),\qquad
\\ \label{RG_eqsT}
\frac{\partial Q_\Gamma^{T}(\eta)}{\partial \ln \Gamma}
&=&\partial_\eta \big[(1 + \eta)Q_\Gamma^{T}(\eta)\big]
+Q_\Gamma^{T}(\eta) [Q_\Gamma^{T}(0)-Q_\Gamma^{I}(0)]
+Q_\Gamma^{I}(0) \theta(\eta -\beta-1)
\int_0^{\eta -\beta-1} \! \! \! \! \! \!\! \! d\eta'Q^{T}_\Gamma(\eta')Q^{T}_\Gamma(\eta - \eta' -\beta- 1).\qquad
\end{eqnarray}
\end{subequations}
\end{widetext}
Here the first term on the right hand side originates from the overall rescaling, the second term corresponds to decimation of the smallest block at cutoff with $\eta=0$, and the last convolution term accounts for the creation of new $I$ or $T$ block of length~$\eta$. 

{\it Fixed point solutions for finite $\alpha$.}---In order to find fixed point distribution we set $ \partial_\Gamma Q_\Gamma^{I,T}(\eta)=0$ in Eqs.~(\ref{RG_eqs}). We find that the value of the fixed point probability distributions $\QIT(\eta)$  at the cutoff can be determined as $I_0 \equiv  \QI(0)= \alpha/({1+\alpha})$ and $T_0 \equiv \QT(0)= 1/({1+\alpha})$~\cite{SOM}. Physically $I_0$ and $T_0$ correspond to the probability to decimate insulating and thermal segments. When $\alpha=1$ we recover the symmetric result of Ref.~\cite{huse} where fixed point was insulating/thermal with probability $1/2$. However, in the limit $\alpha\ll 1$, $ T_0 \to 1$ the fixed point is dominated by insulating regions. Note that the number of blocks scales as $N_\text{tot}\propto 1/\Gamma$ with the cutoff since $I_0+T_0 =1$, so that the total length of the system $\ell_\text{tot}\propto \Gamma N_\text{tot} $ is asymptotically conserved in the RG for $\alpha \beta=1$.  

Using the boundary conditions at $\eta=0$, each equation in system~(\ref{RG_eqs}) can be solved iteratively. 
For the initial region $\eta\in[0,\beta+1]$, the integral term in Eq.~(\ref{RG_eqsT}) vanishes, resulting in a power-law form of $\QT(\eta)$. When $\eta \in [\beta+1,2(\beta+1)]$ one can use the known solution for smaller values of $\eta$ and solve the resulting non-uniform differential equation. Repeating such iterations for both $\QIT(\eta)$, we obtain the fixed point distributions shown in Fig.~\ref{Fig:exponent}(a) for  $\alpha = 1/\beta = 1/10$. The initial power-law region of $\QT(\eta) \sim (1+\eta)^{-(1+T_0-I_0)}$ for $\eta\leq 1+\alpha^{-1}$ is followed by an exponential decay. Since $\alpha$ is small, the power-law region in $\QI(\eta)$ is very short and the distribution can be approximated as $\QI(\eta) = I_0\exp(-I_0\eta)$. 

{\it Critical exponent and fractal dimensions.}---From the fixed point distributions, we obtain the correlation length critical exponent $\nu$ and the fractal dimensions that characterize the fixed point. To extract $\nu$, we consider weak perturbations around the fixed point, parametrized as $Q^{I,T}_\Gamma(\eta) = \QIT(\eta) +  \Gamma^{1/\nu} \fIT (\eta)$.
The critical exponent $\nu$ controls the behavior of the perturbation upon increasing the cutoff, with $\nu>0$ for a relevant perturbation.  We have $\int_0^\infty d\eta\, \fIT(\eta) = 0$ since $\QIT(\eta)$ is normalized to one. Linearizing the RG flows~(\ref{RG_eqs}), we obtain an eigenvalue system of functional equations, $(1/\nu) \fIT(\eta) = \OIT \fIT(\eta)$ where the explicit form of the integro-differential operator $\OIT$ is given in~\cite{SOM}. Solving  this eigenvalue problem, we obtain a single relevant eigenvalue $1/\nu$, which is real and positive and thus sets the critical exponent. The critical exponent $\nu(\alpha)$ takes its minimal value for $\alpha=1$, $\nu(1)\approx 2.50$~\cite{huse} and increases for smaller values of $\alpha$. We note that the increase of $\nu$ when the fixed point becomes more MBL-like qualitatively agrees with other RG approaches~\cite{VHA,PVP} which predict more MBL-like fixed points and suggest $\nu \approx 3.5$. The inverse exponent  $1/\nu$ decays to zero when $\alpha\to 0$. We predict that $\nu^{-1}(\alpha) \approx 1/\ln(1+\alpha^{-1})+{\cal O}(\alpha)$ as $\alpha \to 0$~\cite{SOM}. This is consistent with our results obtained via numerical diagonalization of  $\OIT$, see Fig.~\ref{Fig:exponent}(b).

To quantify the spatial structure of insulating and thermal regions at criticality, we consider their fractal dimensions. For example, the insulating fractal dimension quantifies the scaling of the total length of microscopic insulating segments $\propto \ell^{d_I}$ that are contained in a piece of insulator segment of size $\ell$ after coarse graining, and that were insulating at all RG steps.  As for the critical exponent $\nu$, we obtain the fractal dimensions by solving a linearized eigenvalue problem~\cite{huse,SOM}. The insulating fractal dimension rapidly tends to one as $d_I(\alpha)= 1-  \calO(\alpha^2)$ for small $\alpha$, see Fig.~\ref{Fig:exponent}(c). On the physical grounds fractal thermal inclusions in an MBL region can lead to big rare thermal regions after coarse graining~\cite{Gopa-15,PVP,huse}. In contrast, fractal insulating regions are most likely unphysical as a fractal set of insulating blocks in an otherwise thermal system cannot lead to localization of the full system. We therefore identify $\alpha \to 0$ as a limit susceptible to describe the actual MBLT since in that limit $d_I=1$. In this limit the thermal fractal dimension $d_T(\alpha)$ slowly approaches zero. As we discuss below, this is consistent with the physical picture provided by our RG.

Our analytical results reveal that the RG fixed point becomes increasingly MBL-like as $\alpha$ is decreased. The critical point in the limit $\alpha \to 0$ is localized with probability $T_0=1$, with fractal thermal inclusions. In that limit, the insulating fixed point distribution becomes uniform $Q_{\ast}^{I}(\eta) = \lim_{I_0 \to 0} I_0\exp(-I_0\eta)$, intuitively corresponding to localized blocks of all lengths.  Thermal inclusions are power-law distributed when $\alpha \to 0$, $ Q_T^\ast (\eta) = {(1+\eta)}^{-2}$, consistent with other RGs~\cite{VHA,DVP,TMR}  -- although the exponent differs slightly. Note, that the average length of thermal blocks, $\langle \eta^T \rangle$, diverges logarithmically at the transition, while the typical value $\langle \eta^T \rangle_{\rm typ}$ remains finite. This is consistent with a rare events-driven transition. 
 
\begin{figure}[t]
\begin{center}
\includegraphics[scale=0.8]{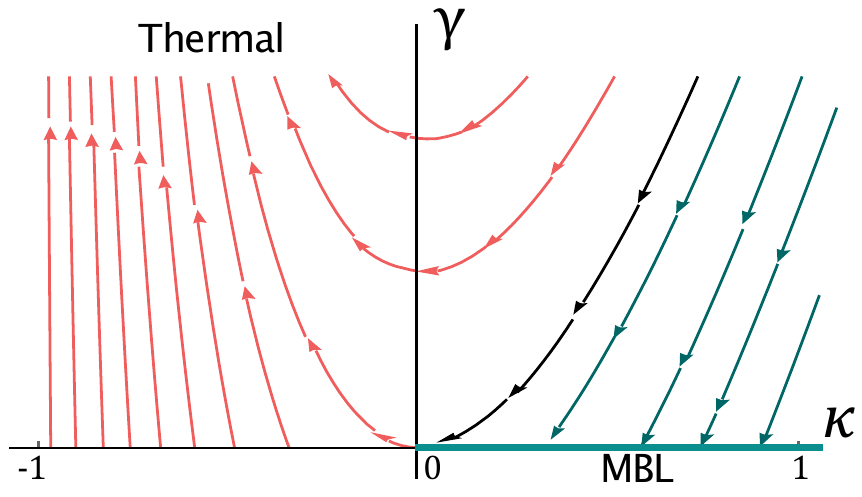}\\
\caption{ \label{Fig:sur}
Two-parameter RG flows in the limit $\alpha\to 0$ has the half-line of stable fixed points $\rt=0$, $\bt>0$  describing a multifractal MBL phase.  For $\rt=0, \bt<0$ the line of fixed points is unstable and gives rise to a flow to strong coupling which corresponds to thermalization.  The black line separates the set of initial conditions that flow to the MBL and thermal phases. 
}
\end{center}
\end{figure}

{\it Two-parameter scaling for the MBLT.}---In the limit $\alpha \to 0$, the eigenfunctions corresponding to the eigenvalue $\nu^{-1} \to 0$ can be determined analytically as $\fI(\eta) = f_{I_0}(1-I_0\eta)e^{-I_0 \eta}$ and $\fT(\eta) = f_{T_0}(1-\ln(1+\eta))/(1+\eta)^2$~\cite{SOM}. Since this perturbation becomes marginal for $\alpha \to 0$, we need to go beyond linear order to analyze the critical behavior. Motivated by the form of the eigenfunctions, we propose the following two-parameter ansatz
\begin{equation}\label{Eq:bInf}
Q^I_\Gamma(\eta) = \rt e^{-\rt \eta},
\quad
Q^T_\Gamma(\eta) = \frac{1+\bt}{(1+\eta)^{2+\bt}},
\end{equation}
where $\rt$ and $\bt$ depend on $\Gamma$ and parametrize deformations of the critical point solution. Both functions are properly normalized provided $\bt>-1$ and $\rt>0$. Moreover, the linear terms in the expansion of Eq.~(\ref{Eq:bInf}) in $\rt, \bt$ are proportional to the critical eigenmodes $\fIT(\eta)$ in the limit $\alpha\to 0$. Plugging this ansatz into Eqs.~(\ref{RG_eqs}), we find that there is an exact line of RG fixed points for $\rt=0$ parametrized by $\bt$. For small $\rt$ the approximate flow equations read:
\begin{equation}\label{Eq:flowInf}
\Gamma \frac{d \rt}{d  \Gamma} = -\rt \bt,
\qquad 
\Gamma \frac{d \bt}{d   \Gamma} = -\rt(1+\bt).
\end{equation}
In contrast to linearized case, the variables $\eta$ and $\Gamma$ do not fully separate, and we neglected a term logarithmic in $\eta$ to get a closed equation for ${d \bt}/{d \Gamma} $~\cite{SOM}. However, the equation for ${d \rt}/{d  \Gamma}$ is accurate for small $\rt$, and both equations correctly predict a line of fixed points for $\rt=0$. 

We conjecture that the two-parameter flow equations~\eqref{Eq:flowInf} correctly capture the critical behavior of the MBLT.  The RG flows are plotted in Fig.~\ref{Fig:sur}, and are equivalent to the celebrated Kosterlitz-Thouless (KT) equations for small $\rt,\bt$~\cite{KT,PhysRevLett.93.150402}. The MBL phase corresponds to a stable line of fixed points with $\rt=0$ and $\bt>0$. This phase has insulating segments of all lengths, with ergodic inclusions distributed algebraically as $\sim \eta^{-(2+\bt_\infty)}$, where $\bt_\infty \geq 0$ parametrizes position on the line of fixed points. While the average length of ergodic regions is finite, the distribution $Q^T_\Gamma(\eta)$ in~(\ref{Eq:bInf}) implies that sufficiently high moments of $\langle(\ell^T)^n\rangle \propto \Gamma^n \langle \eta^n \rangle_{Q^T}  $ with $n\geq 1+\bt_\infty$ diverge, suggesting a multifractal behavior in the MBL phase near the transition~\cite{PhysRevB.93.041424,PhysRevB.93.174202,PhysRevB.96.104201}. The critical point is reached when the (renormalized) exponent $\bt_\infty$ becomes equal to the critical value $\bt_c=0$, which corresponds to the divergence of the average length of thermal inclusions. In our description, the critical point of the MBLT is a smooth continuation of the MBL phase, just like the critical point in the usual KT transition is a superfluid. In the thermal phase, $\rt$ flows to strong coupling corresponding to short insulating regions (in that regime, our RG equations break down), while $\bt$ goes to $-1$ corresponding to infinitely broadly-distributed thermal regions.

The RG trajectories are parametrized by $\rt_\Gamma-\bt_\Gamma+\ln(\bt_\Gamma+1) = C$ where $C>0$ corresponds to the flow into strong coupling thermal phase, while for $C<0$ trajectories flow to the MBL phase. Near the MBLT, we have $C = C_0 (W_c - W) + \dots$ where $W$ corresponds to the bare disorder strength, and $W_c=W$ at the transition. Following usual scaling arguments, the correlation length that sets the crossover to the phases diverges as $\xi \propto \exp(c/\sqrt{\left|W-W_c \right|})$, where $c$ is some non-universal positive constant. This is in sharp contrast with previous approaches that measured a large but finite critical exponent $\nu$,  and that also observed a finite probability to thermalize at criticality drifting with system size~\cite{VHA,PVP,DVP}. Note that the presence of logarithmic finite size corrections characteristic of the KT transitions would make this scaling very hard to observe on finite size systems. The exponent $\kappa_\infty$ on the MBL side of the transition is non-universal, but vanishes as $\kappa_\infty = A \sqrt{W-W_c}$.

{\it Summary and discussion.}---We presented a one-parameter family of RGs that in the limit $\alpha\to0$ provides a sensible description of the MBLT, yet allows for an analytic solution. Our simple two-parameter KT scaling predicts an exponentially diverging correlation length at the transition. The distribution of thermal regions has a power-law form both in the MBL phase, and at the transition. In addition, we recover the absence of fractal insulating regions and a sparse structure of thermal regions that have vanishing fractal dimension. 

These results can be interpreted within a Griffiths picture~\cite{Gopa-15,Griffiths69,agarwal}: in the MBL phase,  thermal inclusions of size $\ell$ require only $\calO(\log \ell)$ independent rare microscopic events with small probability $p$ and therefore occur with algebraic probability $p_T(\ell) \sim p^{\calO( \log \ell)}\sim 1/\ell^{2+\kappa}$. In contrast, rare insulating inclusions of size $\ell$ on the thermal side require  $\calO(\ell)$ rare events and are therefore exponentially distributed $p_I(\ell) \sim {\rm e}^{-\ell/\xi}$, leading to subdiffusive transport properties~\cite{Gopa-15,Reichman15,Luitz-subdiff,Znidaric16,PVP,VHA}. This picture implies that thermal Griffiths inclusions are even sparser than has been previously assumed, since they formally have fractal dimension $d_T=0$. The transition to the thermal side then occurs when $\kappa=\kappa_c=0$. At this point the average size of the thermal inclusions diverges as $\log L$ with system size $L$, and is barely enough to percolate and thermalize the whole system upon coarse graining. It would be very interesting to see if our KT scaling scenario could be tested in other RG schemes~\cite{VHA,PVP,TMR}, numerical studies~\cite{Alet14,Luitz-subdiff,Reichman15,KhemaniEnt} or experiments~\cite{Bloch16G,AbaninRev}.

{\it Acknowledgments.}---RV and MS acknowledge discussions and collaboration with J. Epstein in the early stages of this work. In addition, we acknowledge useful discussions with D. Huse, A. Morningstar, S. Parameswaran,  A.C. Potter, and W. De Roeck. RV also thanks P. Dumitrescu, A.C. Potter and S. Parameswaran for previous collaborations on related topics. This work was supported by the US Department of Energy, Office of Science, Basic Energy Sciences, under Award No. DE-SC0019168 (RV).  AG acknowledges support of the Swiss National Science Foundation.

\bibliography{simpleRG}
\bibliographystyle{apsrev4-1}

\onecolumngrid
\newpage

\section{Supplementary material for ``Analytically solvable renormalization group for the many-body localization transition"}

\begin{quote}
{\small  \label{pg:som}
In this supplementary material we  present detailed derivation of the various results discussed in the main text. We start with discussing interpretation of the microscopic rules and justification of the limit $\alpha\to 0$ in Section~S1. In Section~S2 we obtain the RG flow equations from microscopic RG rules. Next, in Section~S3 we discuss the linearized form of the RG flow equations and show how to extract the critical exponent. Section~S4 presents details of the calculation of fractal dimension along with its asymptotic form. Finally, in Section~S5 we discuss the derivation of RG flow equations in the limit $\alpha\to 0$.
}\\[20pt]
\end{quote}

\section{S1: Interpretation of the microscopic rules of the RG \label{S0}}
Below we discuss the microscopic interpretation of the RG rules, and their limit $\alpha\to 0, \beta \to \infty $ that is considered in the second part of the manuscript. 

As we discuss in the main text, we interpret the length $\ell$ as setting a tunneling rate. In insulating blocks we want to associate a time scale $\tau^I \sim e^{\ell^I/\xi_0}$, with $\xi_0$ being a bare localization length. In contrast, for $T$ blocks we have $\tau^T\sim \ell^T$. Assuming that the new time is set by an inverse product of rates in the $ITI \to I$ decimation process, we obtain 
\begin{equation}\label{Eq:ITI}
\tau_\text{new}^I = \exp\left(\frac{\ell^I_\text{new}}{\xi_0}\right) \sim \exp\left(\frac{\ell^I_{n-1}}{\xi_0}+\ln \ell^T_n + \frac{\ell^I_{n+1}}{\xi_0}\right).
\end{equation} 
Hence the limit $\alpha \to 0$ is motivated by a very small contribution of a thermal block to the transport time through such an insulating region. 

The $TIT \to T$ move is more subtle, since most likely the $I$ block will be a bottleneck in the transport through the $TIT$ region that is now merged into a single block. Hence, even if the length of the $I$ block is much smaller than the length of the neighboring $T$ blocks, the exponential of the length of that block could be much larger, making the dominant time scale $\tau^I_n\sim e^{\ell^I_n/\xi_0}$. Assuming addition of times in the $TIT \to T$ decimation process requires taking
\begin{equation}\label{Eq:TIT}
\tau_\text{new}^T \sim \ell_\text{new}^T \sim \ell_{n-1}^T+\exp\left(\frac{\ell^I_{n}}{\xi_0}\right)+\ell_{n+1}^T,
\end{equation} 
that is different from a simple addition of lengths. Nevertheless,  large $\beta$ in Eq.~(1) in the main text is a way to single out $\ell^I_n$. Combined with the condition $\alpha \beta=1$, this leads to this insulating block being counted with the correct length $\ell_n$ later on in the RG. This choice of rules also ensures that the length of the MBL blocks is asymptotically conserved in the RG.

\section{S2: Calculation of critical point solutions \label{S1}}
Below we present a detailed derivation of the integro-differential Eqs.~(\ref{RG_eqs}) that describe the flow of the probability distributions of thermal and insulating inclusions.  In the coarse-grained description we introduce the density of insulating (thermal) blocks of a certain length $\ell$,  $n^{I, T}_{\Gamma}(\ell)$. For instance,  the number of insulating blocks with length in the interval $[\ell,\ell+d\ell]$ is given by $dN_{\Gamma}^I = n^{I}_{\Gamma}(\ell) d\ell$. In this notation, the total number of blocks of the corresponding type is written as $N^{I, T}_{\Gamma} = \int_{\Gamma}^{\infty}d\ell\, n^{I, T}_{\Gamma}(\ell)$, where we used the fact that cutoff $\Gamma$ coincides with the length of the smallest segment. Next, we define the probability densities as $\rho^{I, T}_{\Gamma}(\ell) = n^{I, T}_{\Gamma}(\ell)/N^{I, T}_{\Gamma}$. In these notations, we can write the change of the density of the blocks from the decimation of blocks with length $\ell \in [\Gamma, \Gamma+\delta \Gamma]$:  
\begin{subequations}
\label{SM_RG_eqs_n}
\begin{align}
n^{I}_{\Gamma +\delta\Gamma}(\ell) &= n^{I}_{\Gamma}(\ell) + n^{T}_{\Gamma}(\Gamma)\delta\Gamma\Big[-2\rho^{I}_{\Gamma}(\ell) + \int_{\Gamma}^{\infty} dl_1\rho^{I}_{\Gamma}(\ell_1)\rho^I_{\Gamma}(\ell-\ell_1-\alpha \Gamma)\Big],\\
n^{T}_{\Gamma +\delta\Gamma}(\ell) &= n^{T}_{\Gamma}(\ell) + n^{I}_{\Gamma}(\Gamma)\delta\Gamma\Big[-2\rho^{T}_{\Gamma}(\ell) + \int_{\Gamma}^{\infty} dl_1\rho^{T}_{\Gamma}(\ell_1)\rho^T_{\Gamma}(\ell-\ell_1-\beta\Gamma)\Big].
\end{align}
\end{subequations}
In these equations $n^{I, T}_{\Gamma}(\Gamma)\delta\Gamma$ is a number of blocks whose length lies in the range $[\Gamma, \Gamma +\delta\Gamma]$, so that they are decimated. The decimation of these blocks leads to creation of new segments with length defined by microscopic RG rules in Eq.~\eqref{Eq:RG_rules}. This is accounted by the integral terms in Eqs.~(\ref{SM_RG_eqs_n}). On the other hand, the negative term in the parenthesis describes the decimation of the blocks of length $\ell$ adjacent to the smallest block.

In order to obtain equations for the probability densities, we notice that total number of insulating (thermal) blocks changes as 
\begin{equation}\label{SM_N}
N^{I, T}_{\Gamma + \delta\Gamma} = N^{I, T}_{\Gamma} - \Big[n^{I,T}_{\Gamma}(\Gamma)+n^{T,I}_{\Gamma}(\Gamma)\Big]\delta\Gamma.
\end{equation}
It reflects the fact that three blocks are merged into one at each RG step. For instance, the total number of insulators is either reduced by decimation of an insulating block at the cut-off, or by the formation of a bigger insulating block in the ITI move. Therefore, the number of the blocks of both types changes in the same manner at each RG step. Assuming that initially they are equal, they stay equal at all the following steps. 
Substituting this into the system~(\ref{SM_RG_eqs_n}) and rewriting the probabilities as functions of $\zeta= \ell-\Gamma$, according to the rules, we get the following equations:
\begin{align}\label{SM_RG_eqs_rho}
&\frac{\partial \rho^{I,T}_{\Gamma}(\zeta)}{\partial\Gamma} = \frac{\partial \rho^{I,T}_{\Gamma}(\zeta)}{\partial\zeta} +\rho^{I,T}_{\Gamma}(\zeta)[\rho^{I,T}_{\Gamma}(0) - \rho^{T,I}_{\Gamma}(0)] + \rho^{T,I}_{\Gamma}(0)\int_0^{\infty} d\zeta_1 \rho^{I,T}_{\Gamma}(\zeta_1)\rho^{I,T}_{\Gamma}(\zeta - \zeta_1 - (1-\alpha[\beta])\Gamma),
\end{align}
Rescaling $\zeta$ as $\eta=\zeta/\Gamma$ and introducing the probability distribution function for rescaled length, $\eta$, as $\rho^{I,T}_{\Gamma}(\ell) = ({1}/{\Gamma})Q^{I,T}(\eta,\Gamma)$, we obtain  Eqs.~(\ref{RG_eqs}) in the main text. 

We note that this two-parameter system of flow equations was studied in the literature for particular values of $\alpha, 
\beta$. Historically, the real-space RG was first developed by Ma, Dasgupta and Hu~\citep{dasgupta_ma, ma_dasgupta_hu} in the context of random 1D Heisenberg model, and further extended by Fisher~\citep{fisher}. In particular, the system of equations~(\ref{RG_eqs}) with $\alpha=\beta=-1$ was used to describe properties of random antiferromagnetic spin chains in Ref.~\citep{fisher}. When $\alpha=\beta=0$  this system coincides with the one considered in Ref.~\citep{rb} to study the competition between the domains of uniform magnetization in one-dimensional scalar systems. Finally, setting $\alpha=\beta=1$, and  $Q^{I}_{\Gamma}(\eta) =  Q^{T}_{\Gamma}(\eta)$ we reproduce the flow equations obtained by Zhang {\it et al.}~\citep{huse}.

In order to find the critical point distributions, we look for stationary solution of the system~(\ref{RG_eqs}), i.e.\ for distributions that do not change with $\Gamma$, ${\partial Q^{I,T}_{\Gamma}(\eta)}/{\partial\Gamma}=0$. 
An exact solution to the above equations is known when $\alpha=\beta=-1$~\citep{fisher} and $\alpha=\beta=0$~\citep{rb}. These solutions were obtained by considering the Laplace transform of the probability distribution function, $\phi_{I,T}(p) = \int_0^{\infty} d\eta e^{-p\eta} Q_\ast^{I,T}(\eta)$. Although the Laplace transform of   Eqs.~(\ref{RG_eqs}) for general values of $\alpha $ and $\beta$ does not allow for analytic solution,  below we demonstrate that it allows to constraint the boundary values of fixed point solutions $I_0 \equiv Q^{I}_\ast(0)$ and $T_0 \equiv Q^T_\ast(0)$.

When written in terms of Laplace image of probability distributions, $\phi_{I,T}(p)$, equations~(\ref{RG_eqs}) with vanishing left hand side read:
\begin{align}\label{Eq:Laplace}
p \phi'_I(p) = p\phi_I(p) - &\phi_I(p) (T_0-I_0) + T_0 \phi^2_I(p) e^{-p(1+\alpha)}-I_0,\\
p \phi'_T(p) = p\phi_T(p) + &\phi_T(p) (T_0-I_0) + I_0 \phi^2_T(p) e^{-p(1+\beta)} -T_0.
\end{align}
These equations are supplemented by the boundary conditions $\phi_I(0) = 1, \phi_T(0) = 1$, that are equivalent to the normalization of the stationary distribution functions $Q_{\ast}^{I,T}(\eta)$.  Using these boundary conditions  and setting $p=0$ in Eqs.~(\ref{Eq:Laplace}) we obtain the following relations:
\begin{align}\label{SM_eqs_for_bc}
(I_0+T_0-1)\phi'_I(0) = T_0(1+\alpha) -1, \quad (I_0+T_0-1)\phi'_T(0) = I_0(1+\beta) -1.
\end{align}
Clearly, setting $I_0 + T_0 = 1$, fixes the boundary terms as $I_0 = \alpha/(1+\alpha)$ and $T_0=1/(1+\alpha)$, restricting the possible values of $\alpha$ and $\beta$: $\alpha\beta=1$. As discussed in the main text, these boundary values are important since they set the probability of system to be insulating/thermal at the fixed point. Moreover, the condition $I_0+T_0=1$ is equivalent to the asymptotic conservation of the total length $\ell_\text{tot} \sim \Gamma N_{\Gamma}$ at the fixed point ($N_{\Gamma}$ is the total number of blocks at cutoff $\Gamma$).  Writing Eq.~\eqref{SM_N} in the differential form, we obtain $dN_{\Gamma}/d\Gamma = -(I_0 + T_0) N_{\Gamma}/\Gamma$.  Thus, provided $I_0+T_0=1$  we have $N_{\Gamma}\sim 1/\Gamma$,  and $\ell_\text{tot}\sim \text{const}$.

\begin{figure}[b]
\begin{center}
\includegraphics[width=0.99\columnwidth]{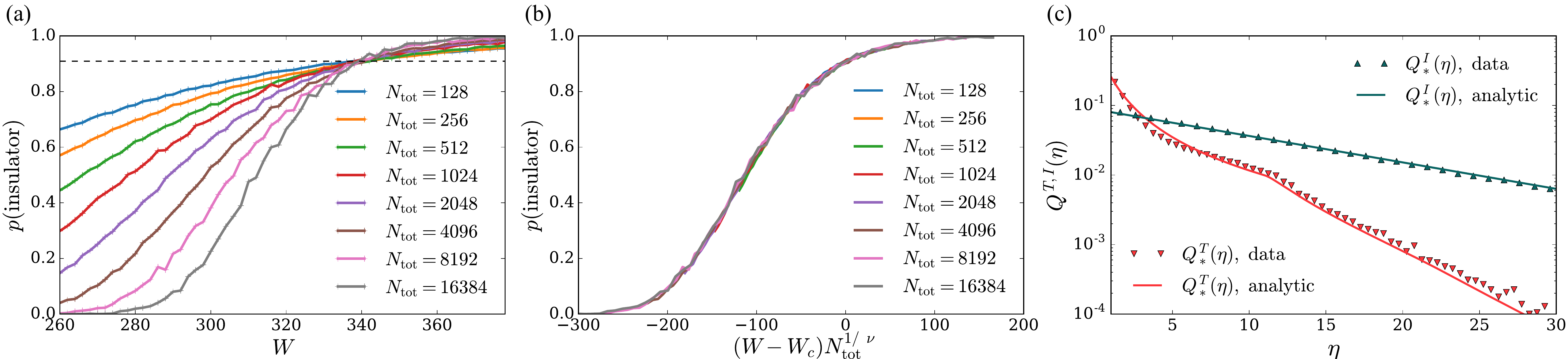}
\caption{ \label{FigS1}
Numerical simulation of RG for $\alpha=1/10$. Left panel shows the probability to have the insulator in the end of RG flow for different number of initial blocks. The dashed line shows the theoretical prediction for the probability to get an insulator, $T_0=10/11$, which coincides with the crossing of curves. Center panel shows the resulting scaling collapse of the curves, where we used the value of exponent $1/\nu = 0.298$ extracted from analytical solutions.
Right panel compares the results of numerical simulation with the analytic predictions for the fixed point distributions.}
\end{center}
\end{figure}

Once we fixed the values of $I_0, T_0$, the system~(\ref{RG_eqs}) separates into two decoupled equations, which can be solved in a piecewise manner. The solutions that we get are in perfect agreement with the numerical simulations of the RG procedure, see Fig.~\ref{FigS1}. In order to numerically simulate the RG flow we start with the $N_\text{tot}$ segments. The thermal/insulating blocks correspond to the even/odd elements of the array. The length of thermal and insulating segments are initially taken from a box distribution of a certain widths $W_T$ and $W_I$. We fix value of $W_T=100$, and use value of $W=W_I$ as a tuning parameter.  Then, the blocks are merged according to the rules so that in the end one is left with the two blocks, the largest of which defines the phase. The transition can be found by varying $W$ and plotting the probability to end in thermal phase as a function of $W$, see Fig.~\ref{FigS1}(a). The resulting finite-size data collapses well when plotted as a function of rescaled disorder strength, when we use the value of critical exponent obtained analytically,  see Fig.~\ref{FigS1}(b).

After approximately determining the value of disorder approximately corresponding to the transition, we run the RG procedure for a system with $N_\text{tot} = 10^6$ blocks initialized at critical value of disorder. When the number of remaining blocks is $N\leq 5\cdot 10^3$ we calculate the properly rescaled distributions of thermal and insulating segments. This procedure is repeated for at least 250 disorder realizations, the resulting average distributions are shown in the right panel of Figure~\ref{FigS1}(c). We observe a perfect agreement between the analytical results for $Q_*^I(\eta)$ and numerical simulation. The agreement for $Q_*^T(\eta)$ is less spectacular, as thermal segments generally have broader distributions, and are expected to be more sensitive to the error in determining the critical value of disorder, $W_c = 337 \pm 0.25$.

While throughout this work we restrict to the solutions on the line $\alpha\beta=1$, we note that the stationary point solutions can be generally found for arbitrary values of $\alpha$ and $\beta$. The additional complication that arises in the most general case is that the initial conditions $I_0$ and $T_0$ need to be found self-consistently to ensure the correct normalization of the solutions. However, they can be found through an iterative procedure. One can start with a guess on the boundary conditions, and solve the equations  Eqs.~(\ref{RG_eqs}) numerically, constructing the solutions in a piecewise manner. Then, we calculate $\phi'_{I,T}(0)$, which is proportional to the average $\eta$, $\phi'_{I,T}(0)=-\int_0^{\infty}d\eta\, \eta Q_{\ast}^{I,T}(\eta)$. Using derivative at the boundary, $\phi'_{I,T}(0)$, we obtain the new values of $I_0$ and $T_0$ from Eq.~\eqref{SM_eqs_for_bc}. Such procedure is repeated until the convergence of boundary values and distribution functions.

\section{S3: Calculation of the correlation length exponent}\label{sec:SM_crit_exp}

\subsection{S3.1: Linearized flow equations}
In order to extract the critical exponent, we need to study the response of the system to small perturbations of fixed point solutions.  We parametrize such perturbation as: 
\begin{equation}
Q^{I}_\Gamma(\eta) = Q_{\ast}^{I}(\eta) + a(\Gamma)\fI(\eta),\quad Q^T_\Gamma(\eta) = Q_{\ast}^T(\eta) + b(\Gamma)\fT(\eta),
\end{equation}
where we assumed that variables $\eta$ and $\Gamma$ can be separated. This condition requires the perturbation to be symmetric with $a(\Gamma)=b(\Gamma)=\Gamma^{1/\nu}$, where $\nu$ is a critical exponent. Substituting the above expression for $Q_\Gamma^{I,T}(\eta)$ into Eqs.~(\ref{RG_eqs}), we obtain the following linearized equations:
\begin{subequations}\label{SM_RG_linearized}
\begin{align}
\nu^{-1} \fI(\eta) &= (1+\eta)\frac{d\fI(\eta)}{d\eta}+ Q_{\ast}^{I}(\eta)[f^{I}(0)-f^{T}(0)] + \fI(\eta)(1+I_0-T_0)\\\nonumber
&+\theta(\eta-1-\alpha)\int_0^{\eta-1-\alpha}d\eta'Q_{\ast}^{I}(\eta')\Big[ 2T_0\fI(\eta-\eta'-1-\alpha)+f^{T}(0)Q_{\ast}^{I}(\eta-\eta'-1-\alpha)\Big],\\
\nu^{-1} \fT(\eta) &= (1+\eta)\frac{d\fT(\eta)}{d\eta} + Q_{\ast}^{T}(\eta)[f^{T}(0)-f^{I}(0)] + \fT(\eta)(1+T_0-I_0)\\\nonumber
&+\theta(\eta-1-\beta)\int_0^{\eta-1-\beta}d\eta'Q_{\ast}^{T}(\eta')\Big[ 2I_0 \fT(\eta-\eta'-1-\beta)+f^{I}(0)Q_{\ast}^{T}(\eta-\eta'-1-\beta)\Big].
\end{align}
\end{subequations}
Thus, finding a relevant critical exponent $\nu$ reduces to solving an eigenvalue problem. Such eigenvalue problem is solved by discretization of  the above integro-differential operators in a sufficiently large region and monitoring the convergence of the results with respect to the region size and discretization step. 

An important subtlety in the calculation of the critical exponent is the careful regularization of the derivative at one of the ends of the interval. Otherwise, one obtains an ill-conditioned matrix with eigenvalues that are extremely sensitive to the size of the interval and discretization step. In this work we use the second order right derivative discretization 
\begin{align}
f'(\eta_n) = \frac{1}{h}\Big[-\frac{1}{2}f(\eta_{n+2})+2f(\eta_{n+1})-\frac{3}{2}f(\eta_n)\Big],
\end{align}
defined on the grid $\{\eta_n\}$ with a total number of points $N$ and a discretization step $h=\eta_{n}-\eta_{n-1}$. Such derivative is well-defined until the last two points of the interval. The derivative at the point $N-1$ is calculated as $f'(\eta_{N-1}) = [f(\eta_{N})-f(\eta_{N-1})]/{h}$. In order to regularize the derivative at the rightmost point, $\eta_N$, we use the asymptotic behavior of $\fIT(\eta)$ at large $\eta$. As the critical solutions both have exponential tails $Q_{\ast}^{I,T}(\eta) \sim e^{-\Lambda_{I,T}\eta}$, Eqs.~(\ref{SM_RG_linearized}) imply that  $\fIT(\eta)\sim \eta e^{-\Lambda_{I,T}\eta}$. Generally, the exponents $\Lambda_{I,T}$ are found numerically, but one can check that in the case of small $\alpha$, $\Lambda_I \simeq I_0$, so that the total fixed point solution for the insulator behaves as $Q_{\ast}^I(\eta) = I_0 e^{-I_0\eta}$. Thus, it allows to regularize the derivative at the rightmost point of the interval as
\begin{equation}
f'(\eta_N) = -\Lambda f(\eta_N) + \frac{f(\eta_N)}{\eta_N}.
\end{equation}
Numerically diagonalizing  the properly discretized operator in Eq.~\eqref{SM_RG_linearized} we obtain a set of its eigenvalues $\{\lambda\}$. Note, that this operator is non-symmetric, thus generally it has a complex valued spectrum.  The inverse critical exponent $\nu^{-1}=\lambda$ is set by  the largest \emph{positive} eigenvalue $\lambda>0$ with correctly normalized eigenfunctions $\int_0^{\infty}d\eta \fIT(\eta)=0$.  We find that for any values of $\alpha$, $\beta$ there always exists a unique eigenvalue satisfying such criteria. In particular, when this operator has more than one positive eigenvalue, the remaining eigenvalues satisfy $\lambda=I_0+T_0$, and hence they cannot be normalized. More specifically, integrating Eqs.~(\ref{SM_RG_linearized}) over $\eta$ results in a condition,
\be
(\lambda - I_0-T_0) \int_0^{\infty}d\eta\, \fIT(\eta)=0.
\ee
This equation implies that the eigenfunctions $\fIT(\eta)$ with $\lambda \neq I_0+T_0$ are automatically normalized, whereas when $\lambda=I_0+T_0$, the eigenfunction may not be normalizable. Numerically we observe that this is indeed the case, as at least one of  the eigenmodes  with eigenvalue  $\lambda=I_0+T_0=1$  never changes sign, and hence cannot be normalized. The same situation was discussed in Ref.~[\onlinecite{huse}], for the case when $\alpha=\beta=1$. 

Numerically implementing the above procedure, we obtain Fig.~\ref{Fig:exponent}b presented in the main text. This figure shows the values of inverse critical exponent for $1/\alpha$ up to $1/\alpha=30$. Calculating the critical exponent at such value of $\alpha$ requires taking  a small discretization step (we observe convergence for $h\leq 0.05$) and an interval of  $\eta\in [0,160]$. The numerical fits to the plot show the slow decrease of $\nu^{-1}$ towards zero upon decreasing $\alpha$.  In order to better understand how the inverse critical exponent behaves for small $\alpha$, below we discuss its analytical asymptotic form. We show that $\nu^{-1}$ vanishes as $1/\ln(1+\alpha^{-1})$, hence the perturbation becomes marginal in the limit of small $\alpha$.

\subsection{S3.2: Asymptotic expression of the critical exponent}
\begin{figure}[tb]
\begin{center}
\includegraphics[scale=0.5]{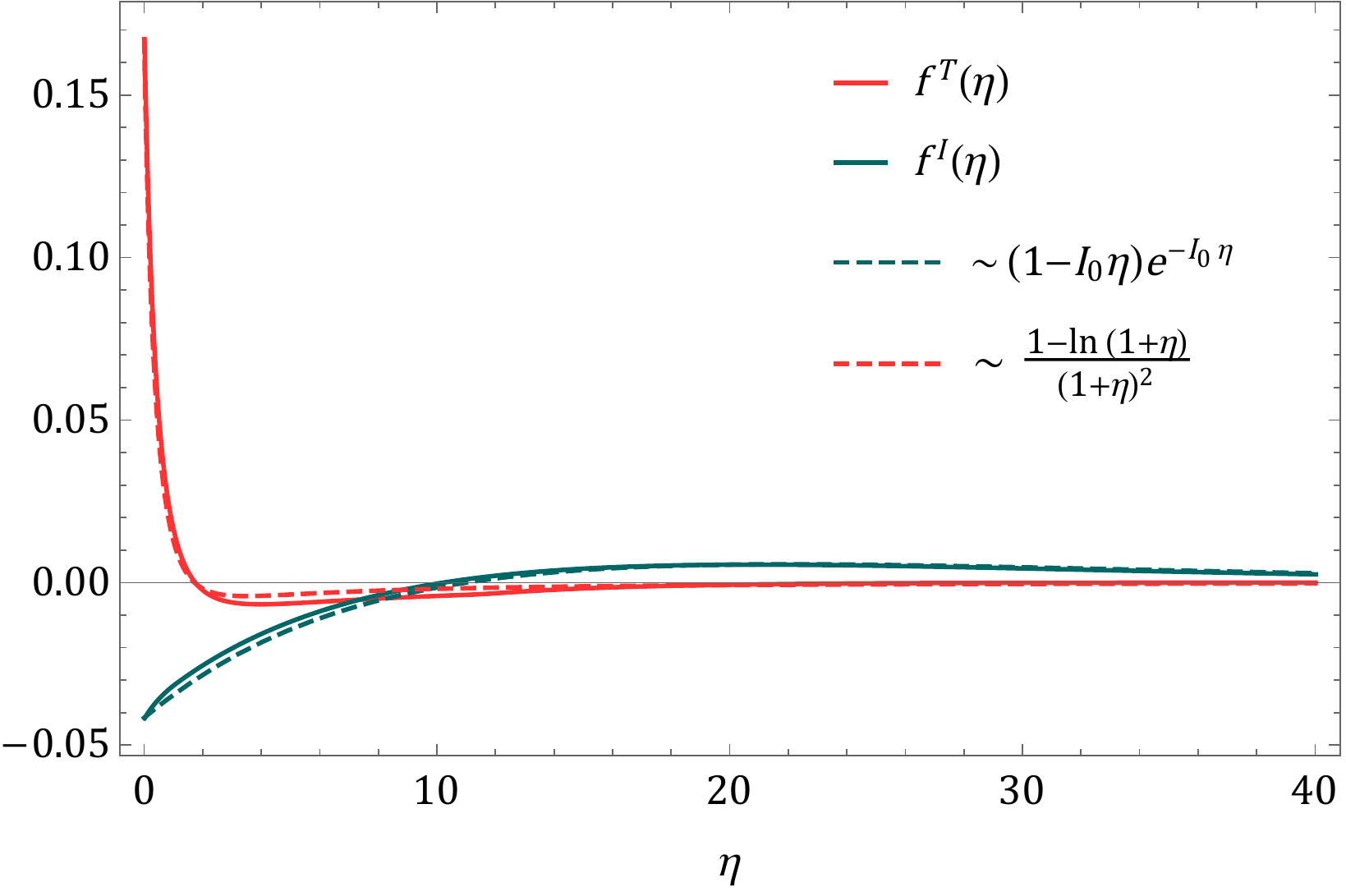}
\caption{ \label{Fig:eigenmodesS}
Eigenmodes $\fIT(\eta)$ for $1/\alpha=10$ are well fitted by our analytical predictions.}
\end{center}
\end{figure}

\begin{figure}[t]
\centering
\includegraphics[scale=0.5]{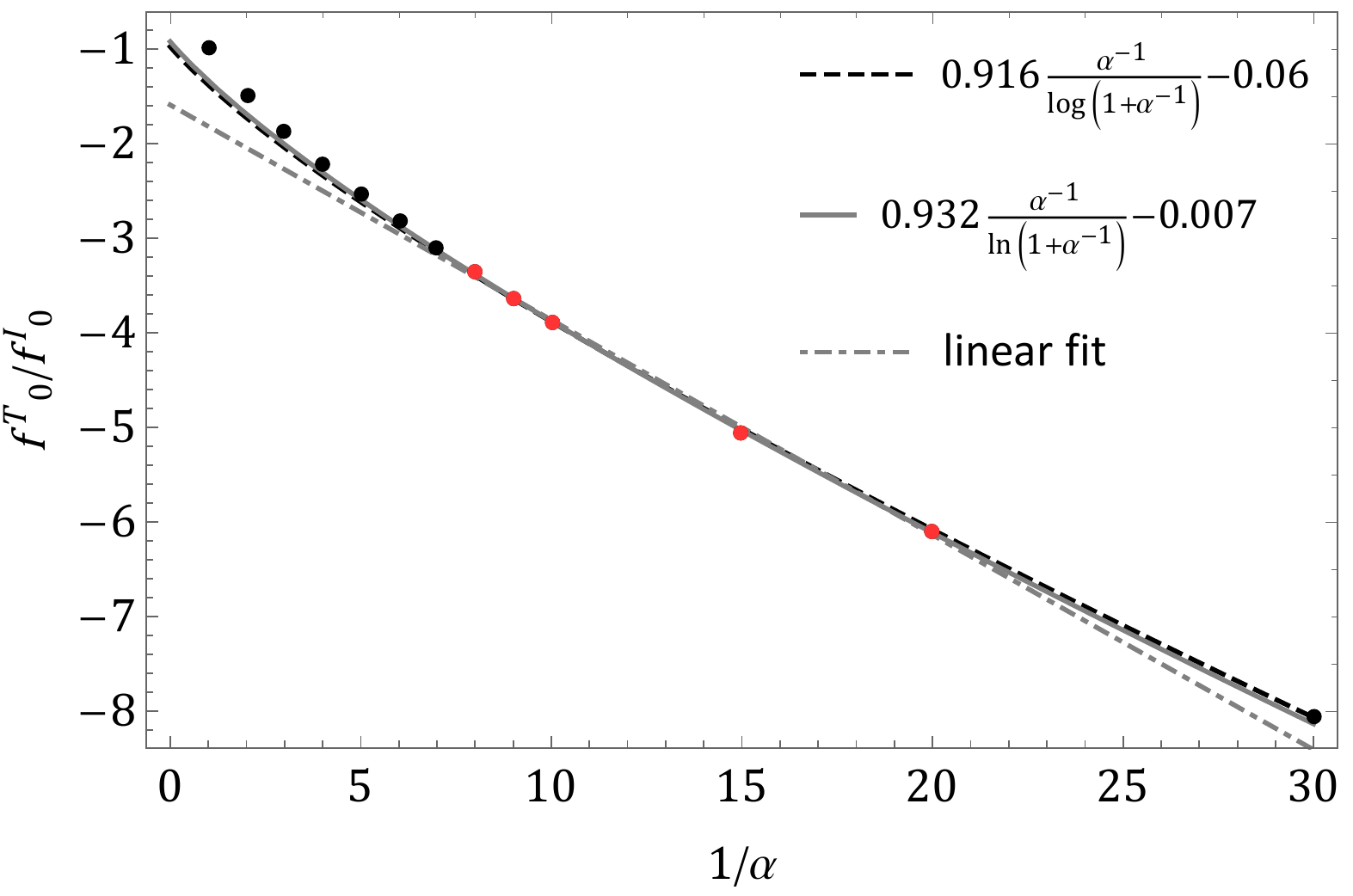} 
\caption{ The ratio of the boundary values $f^{T}(0)/f^{I}(0)$ as a function of $\alpha$. The red points are fitted with a linear curve (dot-dashed) and with $\alpha^{-1}/\ln(1+\alpha^{-1})$ (dashed). Clearly the point corresponding to $1/\alpha=30$ falls well onto $\alpha^{-1}/\ln(1+\alpha^{-1})$ fit. The solid line corresponds to the fit with the last point included. Note, that including the last point only weakly changes the fit parameters.}
\label{Fig:SM_k}
\end{figure}

Let us begin this section by presenting an analytical observation justifying asymptotic behavior  $\nu^{-1} \to 0$ when $\alpha\to0$.  In this limit the critical thermal distribution acquires the form $Q_{\ast}^T(\eta) = 1/(1+\eta)^2$, implying that the integral term in Eq.~\eqref{RG_eqsT} can be neglected. Indeed, such a power-law distribution is properly normalized to unity and solves the Eq.~\eqref{RG_eqsT} without the integral term. Thus, we naturally assume that the integral term can also be ignored in Eqs.~\eqref{SM_RG_linearized} for the eigenmode $f^T(\eta)$. Then, there are generally two possibilities of a finite and zero $\lambda$. If it is finite, the solution is a sum of two power-laws:
\begin{align}\label{fT}
\fT(\eta) = \frac{1}{(1+\eta)^{1+T_0-I_0}}\Big[C_1+ (1+\eta)^{\lambda}(f^{T}_0-C_1)\Big],\quad C_1 = T_0\frac{f^{T}_0-f^{I}_0}{\lambda},
\end{align}
where we denoted the boundary values as $f^T(0)\equiv f^T_0$ and $f^I(0) \equiv f^I_0$. This solution can be smoothly extrapolated to the case of $\lambda=0$.
Indeed, expanding Eq.~\eqref{fT} in $\lambda$ and assuming that $|f^{T}_0/f^{I}_0|\gg 1$ for $\alpha\to 0$ (which we support numerically below, see Fig.\ \ref{Fig:SM_k}), we arrive at
\begin{align}
\fT(\eta) &= f^{T}_0\frac{1-T_0 \ln(1+\eta)}{(1+\eta)^{1+T_0-I_0}} \simeq f^{T}_0 \frac{1-\ln(1+\eta)}{(1+\eta)^2},
\end{align}
which is exactly the solution of Eq.~(\ref{SM_RG_linearized}b) for $\lambda = 0$. Notably, it is well-normalized to zero. The eigenmode for the insulator can be easily found from the fact that the critical point solution is described quite well by $I_0 e^{-I_0\eta}$ in the case of $\alpha\to 0$, i.e. $I_0\to 0$ : $f^I(\eta) = f^{I}_0(1-I_0\eta)e^{-I_0 \eta}$. 
The zero $\lambda$ analytical fit for the thermal and the linear-exponential one for the insulator eigenmodes describe the numerically found solutions quite well already for a case of a finite $1/\alpha$, see Fig.~\ref{Fig:eigenmodesS}.
In order to understand the asymptotic behavior of $\nu^{-1}=\lambda$ for small $\alpha$ we minimize the norm of the residual 
\be
{\cal N}[\lambda] = \int_0^{\infty}d\eta\, \left[\hat{O}_If^I(\eta)-\lambda f^I(\eta)\right]^2,
\ee
where  $\hat{O}_I$ is the integro-differential operator in the r.h.s of Eq.~(\ref{SM_RG_linearized}).  The optimal value of $\lambda$ that minimizes the residual is given by  
\be\label{SM_lambda_scaling}
\lambda(\alpha) = -\alpha\Big[1+k(\alpha)\Big] + \calO\left(\alpha^2\right),\qquad k(\alpha) = \frac{f^{T}_0}{f^{I}_0}.
\ee
Hence, we see that the ratio $k(\alpha)$ between values of $T,I$ eigenmodes at $\eta=0$ sets the asymptotic behavior of the eigenvalue $\lambda(\alpha)$. While we were not able to obtain the analytic asymptotics for $k(\alpha)$, numerical results shown in Fig.~\ref{Fig:SM_k} are well fitted by dependence $k(\alpha)\approx -\alpha^{-1}/\ln(1+\alpha^{-1})$. 
Using such asymptotics for $k(\alpha)$, we obtain:
\begin{equation}\label{Eq:lambda-as}
\lambda(\alpha) = \frac{1}{\ln (1+\alpha^{-1})} + \calO\left(\alpha\right).
\end{equation}
Figure ~\ref{Fig:exponent}b in the main text  shows fit consistent with such asymptotic behavior, although with different prefactors.  

The vanishing inverse critical exponent suggests that in the limit $\alpha\to 0$  another relevant perturbation may emerge. In Section~\ref{S4} below we discuss how one can gain intuition into the form of this perturbation using the eigenmodes $f^{I,T}(\eta)$ in the limit of small $\alpha$.

\section{S4: Calculation of the fractal dimensions \label{S3}}

Before presenting details of calculations, let us start with the discussion of the concept of fractal dimension and its definition.  Consider a thermal block at a particular step of our RG. If this block was never decimated before, we call it ``microscopic'' segment. On the other hand, if this block was created at some stage of RG flow by decimating other blocks, it is natural to ask what is its ``microscopic'' content. The fractal dimension can be used to quantify such internal structure. 

More specifically, the fractal dimension sets how the total length $\ell^T$ of all the microscopic thermal blocks that have been used during the RG to construct the given thermal block of length $\ell$ scales with $\ell$. A scaling 
\begin{equation}\label{fract_dim}
\ell^{T} \sim \ell^{d_{T}},\quad 0< d_T < 1
\end{equation}
is a sign of a fractal structure of the thermal inclusions and $d_T$ is called a fractal dimension. The idea that the thermal rare-regions are fractal was proposed in Ref.~\cite{huse}. Due to unphysical symmetry present in the RG by \textcite{huse}, insulating inclusions would have the same fractal structure that was declared to be unphysical. Indeed, one expects that  rare fractal insulating inclusions should not be able to localize an otherwise typical thermal region, in contrast to the rare thermal inclusions that may be sufficient for thermalization.  Thus, the insulating blocks should not be fractal. Below we discuss the calculation of fractal dimensions $d_{T,I}$ and show that the limit $\alpha\to 0$ provides the expected value of $d_I = 1$ implying absence of fractal insulating regions in this limit.

\subsection{S4.1: Linearized equations for the fractal dimensions }

In order to calculate the fractal dimension we need to keep track of the total length of all microscopic thermal (insulating) segments within a given region. We note that microscopic $T$ ($I$) content of the given region can be calculated using the same  rules from Eq.~\eqref{Eq:RG_rules} but with $\alpha=\beta= 0$. Thus, we introduce additional variable $\chi = {\ell^T}/{\Gamma^{d_T}}$ that corresponds to the dimensionless $\ell^T$ and the joint probability distribution $\rho^{T}_{\Gamma}(\eta, \chi) = {Q^{T}_\Gamma(\eta,\chi)}/{\Gamma^{d_T + 1}}$. Similarly, we introduce the joint distribution for the insulator and use the same notation $\chi = {\ell^I}/{\Gamma^{d_I}}$ as the equations for insulating and thermal fractal dimension are decoupled from each other. The condition for the stationary point of the joint probability distribution function reads:
\begin{subequations}\label{SM_fract_eq}
\begin{align}
(1+\eta)\frac{d}{d\eta}Q_{\ast}^I(\eta,\chi) &+ d_I\frac{d}{d\chi}\Big[\chi Q_{\ast}^I(\eta,\chi)\Big] +  Q_{\ast}^I(\eta,\chi) [1+I_0-T_0]\label{SM_fract_eqI}\\
+
&T_0 \theta(\eta - \alpha-1) \int_0^{\eta - 1-\alpha} d\eta_1\int_0^{\chi}d\chi_1Q_{\ast}^I(\eta_1, \chi_1)Q_{\ast}^I(\eta - \eta_1 - 1-\alpha, \chi - \chi_1) = 0,\nonumber\\
(1+\eta)\frac{d}{d\eta}Q_{\ast}^T(\eta,\chi) &+ d_T\frac{d}{d\chi}\Big[\chi Q_{\ast}^T(\eta,\chi)\Big] +Q_{\ast}^T(\eta,\chi) [1+T_0-I_0]\label{SM_fract_eqT}\\
+
&I_0 \theta(\eta - \beta-1) \int_0^{\eta - 1-\beta} d\eta_1\int_0^{\chi}d\chi_1Q_{\ast}^T(\eta_1, \chi_1)Q_{\ast}^T(\eta - \eta_1 - 1-\beta, \chi - \chi_1) = 0.\nonumber
\end{align}
\end{subequations}
These equations can be easily obtained by calculating the derivative of a joint distribution,
\begin{equation}
\frac{\partial \rho^T_\Gamma(\eta,\chi)}{\partial\Gamma} = -\frac{(d_T + 1)Q^T_{\Gamma}(\eta,\chi)}{\Gamma^{d_T + 2}} + \frac{1}{\Gamma^{d_T + 1}}\left(\frac{\partial Q^T_{\Gamma}(\eta,\chi)}{\partial\Gamma}-\frac{\eta}{\Gamma}\frac{\partial Q^T_{\Gamma}(\eta,\chi)}{\partial\eta}-d_T \frac{\chi}{\Gamma}\frac{\partial Q^T_{\Gamma}(\eta,\chi)}{\partial\chi}\right).
\end{equation}
While this system of equations looks complicated, we are interested only in the average $\ell^T$ and its scaling. Hence, we define the first moment of the joint probability distribution,
\begin{equation}\label{Eq:g-def}
g^{I,T}(\eta) = \int_0^{\infty} d\chi\, \chi Q_{\ast}^{I,T}(\eta,\chi),
\end{equation}
and derive the conditions satisfied by $g^{I,T}(\eta)$ at a stationary point. This is done by multiplying Eqs.~(\ref{SM_fract_eq}) by $\chi$ and integrating over $\chi$. It turns out that after such operation the  integral terms  that are quadratic in joint distribution function reduce to the product of  $g^{I,T}(\eta)$ and the stationary point solution for $\eta$, $Q_{\ast}^{I,T}(\eta)$. For instance, in the case of Eq.~(\ref{SM_fract_eqI}), the integral term is transformed as:
\begin{multline}
\int_0^{\infty}d\chi \chi\int_0^{\chi}d\chi_1Q_{\ast}^I(\eta_1, \chi_1)Q_{\ast}^I(\eta - \eta_1 - 1-\alpha, \chi - \chi_1) = \\
\int_0^{\infty}d\chi \int_0^{\chi}d\chi_1 \chi_1 Q_{\ast}^I(\eta_1, \chi_1)Q_{\ast}^I(\eta - \eta_1 - 1-\alpha, \chi - \chi_1) + \int_0^{\infty}d\chi \int_0^{\chi}d\chi_1 (\chi-\chi_1) Q_{\ast}^I(\eta_1, \chi_1)Q_{\ast}^I(\eta - \eta_1 - 1-\alpha, \chi - \chi_1)\\
\simeq 2 g^I(\eta) Q_{\ast}^I(\eta-\eta_1-1-\alpha).
\end{multline}
In the above transformations we extended the upper limit of the integral from $\chi$ to infinity. This approximation is controlled when joint distributions $Q_{\ast}^{I,T}(\eta, \chi)$ have exponential tails. In this case, the main contribution to the integral comes from the finite region of $\chi$ and the upper limit of integrals can be safely extended to infinity. While we did not check the validity of this assumption numerically, our analytical predictions for the fractal dimensions are in agreement with the numerical simulations. 

After a somewhat lengthy algebra, the equations for the stationary solution for $g^{I,T}(\eta)$  reduce to an eigenvalue problem on $d_{T,I}$:
\begin{subequations}
\begin{align}
 d_I g^I(\eta)=(1+\eta)\frac{d}{d\eta}g^I(\eta) &+ g^I(\eta) [1+I_0-T_0]
+ 2T_0 \theta(\eta - \alpha-1) \int_0^{\eta - 1-\alpha} d\eta_1\, Q_{\ast}^I(\eta_1)g^I(\eta - \eta_1 - 1-\alpha),\label{SM_fract_eq_fin_I}\\
d_T g^T(\eta)=(1+\eta)\frac{d}{d\eta}g^T(\eta) &+ g^T(\eta)[1+T_0-I_0]
+ 2I_0 \theta(\eta - \beta-1) \int_0^{\eta - 1-\beta} d\eta_1 \,  Q_{\ast}^T(\eta_1)g^T(\eta - \eta_1 - 1-\beta).\label{SM_fract_eq_fin_T}
\end{align}
\end{subequations}
In contrast to the equations for the critical exponents, Eq.~\eqref{SM_RG_linearized}, the equations for $d_I$ and $d_T$ are fully decoupled from each other. The numerical calculation of the $d_{I,T}$ is performed by the same method as described in Section~\ref{sec:SM_crit_exp}, including the careful regularization of the derivatives by using an asymptotic form of $g^{I,T}(\eta)\sim \eta e^{-\Lambda_{I,T}\eta}$. 

Note that in the symmetric case $\alpha=\beta=1$, we reproduce the results of Ref.~\cite{huse}. The numerical calculation of fractal dimensions away from the symmetric point but on the line $\alpha\beta = 1$ is shown in Fig.~\ref{Fig:exponent}(c). We observe that $d_I\to 1$ for $\alpha\to 0$. Below we support the numerical dependence of $d_I$ by the analytical derivation of $d_I$ in this limit. Although numerical extrapolation does not allow us to rule out a saturation of $d_T$ to a finite value, we provide a physical justification for its vanishing with $1/\alpha$ following a scaling $1/\ln(1+\alpha^{-1}) $ similar to that of the inverse critical exponent $\nu^{-1}$. As we discuss in the main text, zero fractal dimension corresponds to a logarithmic relation between $\ell^T$ and $\ell$, $
\ell^T\sim \ln \ell$. This allows us to match the Griffiths type arguments and the length distributions of the blocks recovered in the limit $\alpha\to 0$.

\subsection{S4.2: Asymptotic expression for the fractal dimension of insulators}
In Fig.~\ref{Fig:exponent}(c) one easily sees that a fractal dimension $d_I$ rapidly approaches unity when $\alpha$ is decreased. This behavior is consistent with the intuition that decreasing $\alpha$ makes thermal phase ``stronger''. Below we discuss how the asymptotic behavior of $d_I$ with $\alpha$ can be obtained analytically. 

When $\alpha\to 0$ it is natural to search for the solution of Eq.~\eqref{SM_fract_eq_fin_I} in the form similar to that of the eigenmode $f^I(\eta)$:
\begin{equation}\label{Eq:gI}
g^I(\eta) \approx (\eta + 1)e^{-I_0 \eta}.
\end{equation}
To find $d_I$  we minimize the residual between the integro-differential operator $\hat{\mathcal{F}}_I$ given by the right hand side of Eq.~\eqref{SM_fract_eq_fin_I} and eigenvalue, 
\begin{equation}\label{Eq:}
\mathcal{N}(d_I) = \int_0^{\infty} d\eta\, \left[\hat{\mathcal{F}}_I(\eta)g^I(\eta) - d_I g^I(\eta)\right]^2.
\end{equation} 
Calculating this integral using explicit form of $g^I(\eta)$ in Eq.~(\ref{Eq:gI}) we obtain:
\begin{align}
\mathcal{N}(d_I) = (1-d_I)^2 \frac{1+2I_0(1+I_0)}{4I_0^3} - (1-d_I)\frac{3+2I_0}{8 I_0} + \text{const}.
\end{align}
Minimizing this expression over $d_I$ results in the following asymptotic behavior:
\begin{equation}
1 - d_I = \frac{3}{4} \frac{1}{(1+\alpha^{-1})^2} - \frac{1}{2}\frac{1}{(1+\alpha^{-1})^3} + \calO\left(\alpha^4\right),
\end{equation}
which shows a good agreement with the numerical data, see Fig.~\ref{Fig:exponent}c. 

\section{S5: RG flow in the limit $\alpha\to 0$ \label{S4}}

In this section we present additional details of the derivation of RG flow equations in the limit $\alpha\to 0$. As we discussed earlier, in this limit, the critical exponent diverges, $\nu\to \infty$ and perturbation considered earlier becomes marginal. Hence, in order to understand the structure of the RG flow, linearized equations are not enough and we need to go to the next order. This is a challenging problem, since we are effectively dealing with a marginal functional RG flow. Indeed, the ``variable'' that flows with $\Gamma$ in our approach is the pair of distribution functions $Q^{I,T}_\Gamma(\eta)$, and we are seeking to understand the second order expansion of non-linear integro-differential operator acting on $Q^{I,T}_\Gamma(\eta)$. 

In order to make progress we use the intuition provided by the understanding of RG flow when $\alpha$ is small but finite. Below, we show that this intuition can be used to write down two-parameter ansatz for the distribution functions  $Q^{I,T}_\Gamma(\eta)$. This ansatz effectively corresponds to the  \emph{projection} of the functional RG flow onto two relevant directions. As we discuss below, due to the normalization condition obeyed by the distribution functions $Q^{I,T}_\Gamma(\eta)$, such projection can be carried out only approximately. At the same time, in Section~\ref{S4.3} we check that the approximations used in such projection are ``self-consistent'' in that the projected RG flow asymptotically conserves the generalized length defined in the main text.  

\subsection{S5.1: Ansatz for distribution functions}
In order to motivate the ansatz in Eq.~(\ref{Eq:bInf}), we consider the form of fixed points solutions in the limit $\alpha\to 0$. Figure~\ref{Fig:exponent}a in the main text shows that $Q^{T}_\ast(\eta)$ tends to the power-law form $1/(1+\eta)^2$. On the other hand, the distribution of the $I$ segments at the fixed point has an exponential form. Next, we consider the form of the eigenmodes $\fIT(\eta)$ when $\beta$ is large. Figure~\ref{Fig:eigenmodesS} shows that these eigenmodes are well-described by  
\begin{equation}\label{Eq:}
\fI(\eta) =f^{I}_0(1-I_0\eta)e^{-I_0\eta},\quad \fT(\eta) = f^{T}_0\frac{1-\ln(1+\eta)}{(1+\eta)^2}\quad  \text{when}\quad \eta<\alpha^{-1}+1.
\end{equation}
Now we observe that function $Q^T_\ast(\eta)+ \kappa_{\Gamma} \fT(\eta)$  coincides with the perturbative expansion of the function
\begin{align}\label{SM_ansatzT}
Q^T(\eta) = \frac{1+\bt_{\Gamma}}{(1+\eta)^{2+\bt_{\Gamma}}}  = \frac{1}{(1+\eta)^{2}} + \bt_{\Gamma} \frac{1-\ln (1+\eta)}{(1+\eta)^{2+\bt_{\Gamma}}} +\calO(\bt_{\Gamma}^2)
\end{align}
around point $\kappa_{\Gamma}=0$ to the first order in $\kappa_{\Gamma}$. This distribution function is properly normalized for any value of $\bt>-1$, and by construction it coincides with $Q^{T}_\ast(\eta)$  when $\bt=0$. Thus we reproduce the first function from the ansatz used in Eq.~(\ref{Eq:bInf}) in the main text. The second function can be motivated in the similar way: the expansion of $Q^I(\eta)$ from Eq.~(\ref{Eq:bInf}) around point $\gamma=I_0$,
\begin{align}\label{SM_ansatzI}
Q^I_{\Gamma}(\eta) = \gamma e^{-\gamma \eta}  = I_0 e^{-I_0\eta} +(\gamma-I_0) (1-I_0 \eta)e^{-I_0\eta} + \calO\left((\gamma-I_0)^2\right),
\end{align}
coincides with the combination $Q^I_\ast(\eta)+ (\gamma-I_0) \fI(\eta)$ with the exception that the term $(\gamma-I_0) \fI(\eta)$ is of the second order. 
Thus, this ansatz is effectively non-perturbative. In particular, $Q^T_{\Gamma}$ has a general power-law solution when $\gamma = 0$, as then the integral term in Eq.\ \eqref{RG_eqsT} can be neglected. However, below we will see that such non-perturbative form of the ansatz does not allow anymore to separate variables $\Gamma$ and $\eta$. Indeed, in the linearized form the variable separation was achieved by the ansatz $Q^T_\Gamma(\eta) = Q^T_\ast(\eta)+ a_\Gamma \fT(\eta)$, so that the $a_\Gamma$ was the only $\Gamma$-dependent parameter. Such distribution function can be normalized for any $a_\Gamma$ provided that $Q^T_\ast(\eta)$ is normalized to one, and $\int d\eta\, \fT(\eta) = 0$. In the present case, Eq.~(\ref{SM_ansatzT}) is still normalized for any value of $\kappa>-1$. Yet, obviously it cannot be represented as a product of a two functions that depend only on $\eta$ and $\Gamma$ respectively.  Importantly, this ansatz captures an exact line of fixed points of the functional RG flows  parametrized by $\kappa$ and $\gamma \to 0$ . 

\subsection{S5.2: Derivation of the RG flow}

After motivating the ansatz used in the main text, we discuss the derivation of the RG flow equations~(\ref{Eq:flowInf}). In order to derive these equations, we insert the expressions for $Q^{T,I}_{\Gamma}$ from Eqs.~(\ref{SM_ansatzT})-(\ref{SM_ansatzI}) into the flow equations Eqs.~(\ref{RG_eqs}).  Note that we assume that $Q^{T,I}_{\Gamma}$ depend on the cutoff $\Gamma$ only via couplings $\bt_{\Gamma}$ and $\gamma_{\Gamma}$ and keep terms up to second order in them. Ignoring the integral in the equation for $Q^T_{\Gamma}$ in the limit $\alpha\to 0$ leads to the following equations
\begin{subequations}\label{SM_reduced_RG_eqs}
\begin{align}\label{Eq:gm}
\Gamma \frac{\partial \gamma}{\partial \Gamma} &= -\gamma \kappa - \gamma^2 (1+\kappa) +  \gamma \left( \gamma\kappa + \Gamma  \frac{\partial \gamma}{\partial \Gamma}\right)\eta,\\ \label{Eq:kp}
\Gamma \frac{\partial \kappa}{\partial \Gamma}& \left[1-(1+\kappa)\log(1+\eta)\right] = -\gamma(1+\kappa).
\end{align}
\end{subequations}
The first term in the equation~(\ref{Eq:gm}) arises from the term linear in $Q^I_{\Gamma}(\eta)$ in Eq.~\eqref{RG_eqsI}, while the rest comes from the integral. In the vicinity of the transition we may neglect the second order terms in $\gamma$, as $\gamma\to 0$. After this we observe that $\eta$-dependent terms drop out from Eq.~(\ref{Eq:gm})  as its prefactor vanishes up to $\calO(\gamma^2)$. The r.h.s. of the equation on $\kappa$ is fully defined by $-Q^I_{\Gamma}(0)Q^T_{\Gamma}(\eta)$ in Eq.~\eqref{RG_eqsT}.
Although the equation on $\kappa$ does not close, the variable $\eta$ enters it only through the logarithm. Neglecting it in the range of small $|\kappa|<1$ brings the final flow equations,
\begin{equation}\label{SM_Eq:flow}
\Gamma \frac{\partial \gamma}{\partial \Gamma} = -\gamma \kappa,\quad \Gamma \frac{\partial \kappa}{\partial \Gamma} = -\gamma(1+\kappa),
\end{equation}
that coincide with Eq.~(\ref{Eq:flowInf}) given in the main text. Despite the approximations made in deriving these equations, we note  that the prediction for the line of the fixed points, $\gamma=0,\kappa>0$, does not depend on these approximations and holds for a general system of equations~\eqref{RG_eqsI} as soon as the integral term can be neglected  in the limit $\alpha \to 0$.

\subsection{S5.3: Asymptotic conservation of the generalized length \label{S4.3}}
The structure of the RG flow~(\ref{SM_Eq:flow}) and the physics behind it are discussed in the main text. Here we would like to perform a self-consistency check of these equations. In the main text we discussed the conservation of the generalized total length by the microscopic RG rules, $\ell_\text{tot} = \sum_n (\alpha\ell^T_n +  \ell^I_n)$. In the limit $\alpha\to 0$, this length reduces to the total length of insulating segments, $\ell^I_\text{tot} = \sum_n   \ell^I_n$. In order to check the conservation of this length in the RG flow, we note that it can be expressed as 
\begin{equation}\label{Eq:litot}
\ell^I_\text{tot} = \frac12 N_{\Gamma}\Gamma (1+\langle\eta\rangle_I),
\end{equation}
where we used the definition of $\eta$ to obtain the average physical length of the insulating block. This average length of insulating segments is multiplied by the total number of insulating blocks, $N_\Gamma/2$, which is given by the half of the total amount of blocks. At each RG step either a thermal or an insulating block is decimated, resulting in a simple relation
\begin{align}
\frac{dN_{\Gamma}}{d\Gamma} = -\left(\rho^T_{\Gamma}(0)+\rho^I_{\Gamma}(0)\right) N_{\Gamma} =-\frac{Q_\Gamma^T(0)+Q^I_\Gamma(0)}{\Gamma}N_{\Gamma}.
\end{align}
Using the values of  $Q_\Gamma^{T,I}(0)$ from Eqs.~(\ref{SM_ansatzT})-(\ref{SM_ansatzI}) and integrating this equation,  we obtain 
\begin{equation}\label{Eq:NG}
 N_{\Gamma} = \frac{C}{\Gamma^{1+\kappa +\gamma}},
\end{equation}
where $C$ is a constant set by initial conditions. Using the expression for $ N_{\Gamma}$ along with average $\langle\eta\rangle_I = 1/\gamma$,  we obtain for $\ell^I_\text{tot}$:
\begin{equation}\label{Eq:litot2}
\ell^I_\text{tot} = \frac{C}{2} \frac{1+\gamma^{-1}}{\Gamma^{\kappa +\gamma}} .
\end{equation}
The conservation of $\ell^I_\text{tot}$ in the RG flow implies that this expression should not change with $\Gamma$.  For the case of small  $\gamma$ and $\kappa>0$, such condition implies following dependence of $\gamma$ on $\Gamma$, $\gamma \sim \Gamma^{-\kappa}$. However, exactly the same dependence follows from the first equation in system~(\ref{SM_Eq:flow}) if we assume that $\kappa$ is constant. More formally, we can differentiate the relation  $\ln \gamma  = -\kappa\ln\Gamma +\text{const}$ with respect to $\Gamma$ and obtain:
\begin{equation}
\frac{d\gamma}{d\Gamma} = -\gamma\left(\frac{\kappa}{\Gamma} -\ln\Gamma \frac{d\kappa}{d\Gamma}\right).
\end{equation}
According to Eq.~(\ref{SM_Eq:flow}), the  derivative of $\kappa$ is proportional to $\gamma$, so the second term in parenthesis is second order in $\gamma$ and thus it can be neglected. The resulting equation coincides with  the flow of $\gamma$ in Eq.~\eqref{SM_Eq:flow}. Therefore, we conclude that despite the approximations made in deriving the RG flow equations~(\ref{SM_Eq:flow}), the neglected terms are asymptotically not important and the conservation of $\ell^I_\text{tot} $ is recovered from RG flow equations in the limit of $\gamma$ being small.

\end{document}